\documentclass[11pt]{article}
\usepackage{acl2015}
\usepackage{times}
\usepackage{url}
\usepackage{latexsym}
\usepackage{amssymb}
\usepackage{amsmath}
\usepackage{graphicx}
\usepackage{enumitem}

\usepackage{bm}
\newcommand{\vect}[1]{\bm{#1}}
\newcommand{\matr}[1]{\bm{#1}}

\newcommand{\vc}[0]{\vect{c}}

\newcommand{\vx}[0]{\vect{x}}
\newcommand{\vz}[0]{\vect{z}}

\newcommand{\vy}[0]{\vect{y}}

\newcommand{\mW}[0]{\matr{W}}

\newcommand{\mV}[0]{\matr{V}}

\title{Transformer for Emotion Recognition}

\author{Jean-Benoit Delbrouck \\
  TCTS Lab, University of Mons, Belgium\\
  {\tt   jean-benoit.delbrouck@umons.ac.be}}

\date{}

\begin{document}
\maketitle
\begin{abstract}
This paper describes the UMONS solution for the OMG-Emotion Challenge\footnote{\url{https://www2.informatik.uni-hamburg.de/wtm/OMG-EmotionChallenge/}}. We explore a context-dependent architecture where the arousal and valence of an utterance are predicted according to its surrounding context (i.e. the preceding and following utterances of the video). We report an improvement when taking into account context for both unimodal and multimodal predictions. Our code is made publicly available\footnote{\url{https://github.com/jbdel/OMG_UMONS_submission/}}.
\end{abstract}

\section{Data}

The organizers specially collected and annotated a One-Minute Gradual-Emotional Behavior dataset (OMG-Emotion dataset) for the challenge. The dataset is composed of Youtube videos chosen through keywords based on long-term emotional behaviors such as "monologues", "auditions", "dialogues" and "emotional scenes". An annotator has to watch a whole video in a sequence so that he takes into consideration the contextual information before annotating the arousal and valence for each utterance of a video. The dataset provided by the organizers contains a train split of 231 videos composed of 2442 utterances and validation split of 60 videos composed of 617 utterances. For each utterance, the gold arousal and valence level is given.

\section{Architecture}

Because context is taken into account during annotation, we propose a context-dependent  architecture \cite{poria2017context}  where the arousal and valence of an utterance is predicted according to the surrounding context. Our model consists of three successive stages:

\begin{itemize}
\item A context-independent unimodal stage to extract linguistic, visual and acoustic features per utterance
\item A context-dependent unimodal stage to extract linguistic, visual and acoustic features per video
\item A context-dependent multimodal stage to make a final prediction per video

\end{itemize}

\subsection{Context-independent Unimodal stage}

Firstly, the unimodal features are extracted from each utterance separately. We use a mean square error as loss function :
$$\mathcal{L}_{mse} = \frac{1}{N}L^2_{p=2} (\vx, \vy) = \frac{1}{N} \sum_{i=1}^{N} (x_i - y_i)^2$$
where $N$ is the number of utterances predicted, $\vx$ the prediction vector for arousal or valence and $y_i$ the ground truth vector.

Below, we explain the linguistic, visual and acoustic feature extraction methods.

\subsubsection{Convolutional Neural Networks for Sentences}

For each utterance, a transcription is given as a written sentence. We train a simple CNN with
one layer of convolution \cite{kim2014convolutional} on top of word vectors obtained from an unsupervised neural language model \cite{mikolov2013distributed}. More precisely, we represent an utterance (here, a sentence) as a sequence of $k$-dimensional word2vec vectors concatenated. Each sentence is wrapped to a window of 50 words which serves as the input to the CNN. Our model has one convolutional layer of three kernels of size 3, 4 and 2 with 30, 30 and 60 feature maps respectively. We then apply a max-overtime
pooling operation over the feature map and capture the most important feature, one with the highest value, for each feature map. Each kernel and max-pooling operation are interleaved with ReLu activation function. Finally, a fully connected network layer $\text{FC}_{out}$ of size $[120 \rightarrow 2]$ predicts both arousal and valence of the utterance. We extract the 120-dimensional features of an utterance before the $\text{FC}_{out}$ operation.

\subsubsection{3D-CNN for visual input}

In this section, we explain how we extract features of each utterance's video with a 3D-CNN \cite{ji20133d}. A video is a sequence of frames of size $W\times H \times 3$. The 3D convolution is achieved by convolving a 3D-kernel to the cube formed by stacking multiple successive video frames together. By this construction, the feature maps in the convolution layer is connected to multiple frames in the previous layer and therefore is able to capture the temporal information. In our experiments, we sample 32 frames of size $32 \times 32$ per video, equally inter-spaced, so that each video in the dataset $\in \mathbb{R}^{32 \times 32 \times 32 \times 3}$. Our CNN consists of 2 convolutional layers of 32 filters of size $5 \times 5 \times 5$. Each layer is followed by two max-pooling layers of size $4 \times 4 \times 4$ and $3 \times 3 \times 3$ respectively. Afterwards, two fully connected network layers $\text{FC}_{out_1}$  $[864 \rightarrow 128]$ and  $\text{FC}_{out_2}$  $[128 \rightarrow 2]$ map the CNN outputs to a predicted arousal and valence level.  We extract the 128-dimensional features of an utterance before the $\text{FC}_{out_2}$ operation.

\subsubsection{OpenSmile for audio input}

For every utterance's video, we sample a Waveform Audio file at 16 KHz frequency and use OpenSmile \cite{eyben2010opensmile} to extract 6373 features from the IS13-ComParE configuration file. To reduce the number, we only select the $k$-best features based on univariate statistical regression tests where arousal and valence levels are the targets. We pick $k=80$ for both arousal and valence tests and merge features indexes together. We ended up with 121 unique features per utterances.

\subsection{Context-dependent Unimodal stage}
In this section, we stack the utterances video-wise for each modality. Lets consider a modality $m$ of utterance feature size $k$, a video $\mV_i$ is the sequence of utterances vectors $(\vx_1, \vx_2, \hdots, \vx_T)_i$ where $\vx_j \in \mathbb{R}^k$ and $T$ is the number of utterances in $\mV_i$. We now have a set of modality videos $\mathcal{V}_m = (\mV_1, \mV_2, \hdots, \mV_N)_m$ where $N$ is number of video in the dataset.\\

In previous similar work \cite{poria2017context}, the video matrice $\mV_i$ was the input of a bi-directional LSTM network to capture previous and following context. We argue that, especially if the video has many utterances, the context might be incomplete or inaccurate for a specific utterance. We tackle the problem by using self-attention (sometimes called intra-attention). This attention mechanism relates different positions of a single sequence in order to compute a representation of the sequence and has been successfully used in a variety of tasks \cite{parikh2016decomposable,lin2017structured,vaswani2017attention}. More specifically, we use the "transformer" encoder with multi-head self-attention to compute our context-dependent unimodal video features.

	\begin{figure}[h!]
		\centering
		\includegraphics[scale=0.40]{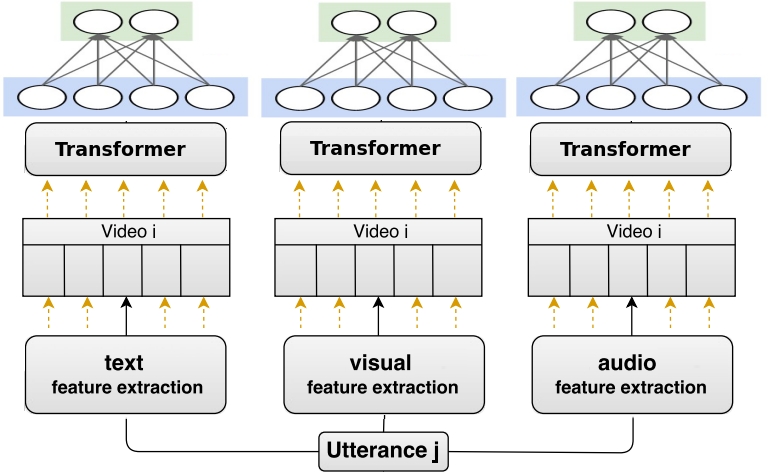}
		\caption{Overview of the Context-dependent Unimodal stage. Each utterance's arousal and valence level are predicted through a whole video}
	\end{figure}

\subsubsection{Transformer encoder}
The encoder is composed of a stack of N identical blocks. Each block has two
layers. The first layer is a multi-head self-attention mechanism, and the second is a fully connected feed-forward network. Each layer is followed by a normalization layer and employs a residual connection. The output of each layer can be rewritten as the following
$$\text{LayerNorm}(x + \text{layer}(x))$$
where layer(x) is the function implemented by the layer
itself (multi-head attention or feed forward).

\subsubsection{Multi-Head attention}

Let $d_k$ be the queries and keys dimension and $d_v$ the values dimension, the attention function is the dot products of the query with all keys, divide each by $\sqrt[]{d_k}$, and apply a softmax function to obtain the weights on the values :

$$\text{Attention}(Q, K, V) = \text{softmax}(\frac{QK^T}{\sqrt[]{d_k}})V$$

Authors found it beneficial to linearly project the queries, keys and values $h$ times with different learned linear projections to $d_k$, $d_k$ and $d_v$ dimensions. The output of the multi-head attention is the concatenation of the $h$ number of $d_v$ values. \\

We pick $d_k = d_v = 64, h = 8, N = 2$.

\subsubsection{Dense output layer}

The output of each utterance's transformer goes through a last fully connected layer $\text{FC}_{out}$ of size $[m, 2]$ to predict both arousal and valence level. Because we make our prediction per video, we propose to include the concordance correlation coefficient in our loss function. We define $\mathcal{L}_\text{ccc} = 1-p_c$ where 

$$p_c = \frac{2\sigma^2_{xy}}{\sigma^2_x + \sigma^2_y + (\mu_x - \mu_y)^2}$$

We now want to minimize 

$$\mathcal{L}_{\text{total}} = \mathcal{L}_{\text{mse}}+0.25 \times \mathcal{L}_{\text{ccc}}$$
for both arousal and valence value. In addition to lead to better results, we found it to give the model more stability between evaluation and reproducibility between runs.

\section{Context-dependent Multimodal stage}
This section is similar to the previous section, except that we now have only one set of video $\mathcal{V} = (\mV_1, \mV_2, \hdots, \mV_N)$ where each video $\mV_i$ is composed of multimodal utterances $\vx_j = (\vx_{\text{linguistic}}, \vx_{\text{visual}}, \vx_{\text{audio}})$. In our experiments, we tried two types of fusion.

\begin{enumerate}[label=\textbf{\Alph*}]
\item \textbf{Concatenation} \\
We simply concatenate each modality utterance-wise. The utterance $\vx_j$ can be rewritten $\vx_j = \vx_{\text{linguistic}} \oplus \vx_{\text{visual}} \oplus \vx_{\text{audio}}$ where $ \oplus$ denotes concatenation.
\item \textbf{Multimodal Compact Bilinear Pooling} \\
We would like each feature of each modality to combine with each others. We would learn a model $\mW$ (here linear), i.e. $\vc = \mW [ [ \vx \otimes \vy ] \otimes \vz]$ where $\otimes$ is the outer-product operation and $[ ]$ denotes linearizing the matrix in a vector. In our experiments, our modality feature size are 120, 128 and 121. If we want $c \in \mathbb{R}^{512}$, $\mW$ would have 951 millions parameters. A multimodal compact bilinear pooling model \cite{fukui2016multimodal} can be learned  by relying on the Count Sketch projection function \cite{charikar2002finding}  to project the outer product to a lower dimensional space, which reduces the number of parameters in $\mW$.

\end{enumerate}

\section{Results}

We report our preliminary results in term of the concordance correlation coefficient metric. \\

\begin{tabular}{lc}
			\multicolumn{1}{c}{\bf Model}  &\multicolumn{1}{c}{\bf 				Mean CCC}
			\\ \hline \\
            \textbf{Monomodal feature extraction} \\
            Text - CNN & 0.165\\
            Audio - OpenSmile & 0.150\\
            Video - 3DCNN	 & 0.186\\
            \textbf{Contextual monomodal} \\
			Text & 0.220\\
            Audio & 0.223\\
            Video & 0.227\\
           \textbf{Contextual multimodal} \\
 			T + A + V & 0.274\\
			T + A + V + CBP & 0.301

\end{tabular}

\bibliographystyle{acl}
\bibliography{acl2015}

\end{document}